\RequirePackage{fix-cm}
\documentclass{svjour3}                     
\smartqed  
\usepackage{graphicx}
\usepackage{color}
\usepackage[colorlinks,
            citecolor=blue,
            linkcolor=blue,
            hypertex,
            breaklinks=true
            ]{hyperref}

\journalname{Int J Theor Phys}
\begin{document}

\title{Meson effect in the proto neutron star PSR J0348+0432
}

\titlerunning{Meson effect in the proto neutron star PSR J0348+0432}        

\author{Xian-Feng Zhao$^{1,2}$   
}


\institute{Xian-Feng Zhao \at
              $^{1}$ School of Sciences, Southwest Petroleum University, Chengdu, 610500 China\\
              $^{2}$ School of Electronic and Electrical Engineering, Chuzhou University, Chuzhou 239000, China\\
              \email{zhaopioneer.student@sina.com}           
}

\date{Received: \today / Accepted: date}

\maketitle

\begin{abstract}
The effect of mesons $f_{0}(975)$ (named as $f$), $\phi(1020)$ (named as $\phi$) and $\delta$ on the moment of inertia of the PNS PSR J0348+0432 is examined in the framework of the relativistic mean field theory considering the baryon octet. It is found that the energy density $\varepsilon$ and pressure $p$ will increase considering the mesons $\delta$ whereas will decrease as the mesons $f$ and $\phi$ being considered. When the mesons $f, \phi$ and $\delta$ are considered, the energy density and pressure will all decrease. It is also found that the contribution of mesons $f$, $\phi$ and $\delta$ to the central energy density is only the central energy density's 0.06$\sim$0.6\% whereas the contribution of mesons $f$, $\phi$ and $\delta$ to the central pressure is the central pressure's 4$\sim$7\%. For the radius, it will decrease when the contributions of mesons $f$, $\phi$ and $\delta$ are considered. The moment of inertia $I$ will increase considering the mesons $\delta$ whereas will decrease as the mesons $f$ and $\phi$ being considered. When the mesons $f$, $\phi$ and $\delta$ are all considered, the moment of inertia will decrease. It is found that the contribution of mesons $f$ and $\phi$ to moment of inertia is 4$\sim$9 times larger than that of mesons $\delta$. Our results show that the mesons $f$, $ \phi$ and $\delta$ contribute to the moment of inertia's 2$\sim$5\%.

\keywords{moment of inertia \and relativistic mean field thoery \and neutron star}
\PACS{26.60.Kp \and 21.65.Mn }
\end{abstract}

\section{Introduction}
The moment of inertia is an important physical quantity for the rotation of a neutron star (NS). Starting from the theory of general relativity, the formula for calculating the moment of inertia of a slowly rotating NS with spherical symmetry was derived by Hartle et al in 1967 and 1968~\cite{Hartle67,Hartle68}. This formula reflects the complex relation between moment of inertia and mass and radius.

The moment of inertia of the NS PSR J0737-3039A with the mass of $M$=1.338 M$_{\odot}$ was studied with five different kinds of equations of state by Bejger et al in 2005~\cite{Bejger05}. The moment of inertia they determined is in the range $0.98\times10^{45}\sim1.72\times10^{45}$ g cm$^{2}$. It was further identified as $I$=1.3$\times$10$^{45}$g cm$^{2}$ by Raithel et al in 2016~\cite{Raithel16}. In their work, they calculated the radii and moments of inertia of the NS PSR J0737-3039A by 41 different equations of state.

Last several years, two massive NSs have been successively discovered. In 2010, the massive NS PSR J1614-2230 with mass of 1.97$^{+0.04}_{-0.04}$ M$_\odot$ was found by Demorest et al~\cite{Dem10}. Two years later, another more larger mass NS PSR J0348+0432 with the mass of M=$2.01\pm0.04$ M$_{\odot}$ was discovered by Antoniadis et al~\cite{Anton13}. Thereafter, many theoretical studies have been carried out on them~\cite{{Mass12},{Masu12},{Malli12},{Whitt12},{Jiang12},{Weiss12},{Cham12},{Kata12},{Weis12},{Bedn12}}.

For the calculations of the NS matter, the relativistic mean field (RMF) theory is a better method~\cite{Zhou16}.
In its early application, only mesons $\sigma$, $\omega$ and $\rho$, by which only the interactions between nucleons can be described, were considered~\cite{Glendenning85}.  Later, mesons $f_{0}(975)$ (named as $f$) and $\phi(1020)$ (named as $\phi$) were introduced to describe the interaction between hyperons~\cite{Schaffner94}.
In order to describe the isospin asymmetry properties of NS matter, mesons $\delta$ also needs to be introduced~\cite{Menezes04}. We call this the $\sigma\omega\rho f \phi \delta$ model.

The proto neutron star (PNS), which emits energy by neutrino radiation, is a very important celestial body for the understanding of the stellar evolution~\cite{Burrows86}. Will the introduction of the new mesons have a great impact on the properties of the PNS? We are interested in this issue.

In this paper, the effect of mesons $f$s, $\phi$s and $\delta$s on the moment of inertia of the PNS PSR J0348+0432 is examined in the framework of the RMF theory considering the baryon octet.

\section{The RMF theory of a PNS}
The Lagrangian density of hadron matter containing mesons $f$, $\phi$ and $\delta$ reads as follows~\cite{Schaffner94,Hofmann01,Kubis97,Glen97}
\begin{eqnarray}
\mathcal{L}&=&
\sum_{B}\overline{\Psi}_{B}(i\gamma_{\mu}\partial^{\mu}-{m}_{B}+g_{\sigma B}\sigma-g_{\omega B}\gamma_{\mu}\omega^{\mu}
\nonumber\\
&&-\frac{1}{2}g_{\rho B}\gamma_{\mu}\tau\cdot\rho^{\mu})\Psi_{B}+\frac{1}{2}\left(\partial_{\mu}\sigma\partial^{\mu}\sigma-m_{\sigma}^{2}\sigma^{2}\right)
\nonumber\\
&&-\frac{1}{4}\omega_{\mu \nu}\omega^{\mu \nu}+\frac{1}{2}m_{\omega}^{2}\omega_{\mu}\omega^{\mu}-\frac{1}{4}\rho_{\mu \nu}\cdot\rho^{\mu \nu}
\nonumber\\
&&+\frac{1}{2}m_{\rho}^{2}\rho_{\mu}\cdot\rho^\mu-\frac{1}{3}g_{2}\sigma^{3}-\frac{1}{4}g_{3}\sigma^{4}
\nonumber\\
&&+\sum_{\lambda=e,\mu}\overline{\Psi}_{\lambda}\left(i\gamma_{\mu}\partial^{\mu}
-m_{\lambda}\right)\Psi_{\lambda}
\nonumber\\
&&+\mathcal{L}^{YY}+\mathcal{L}^{\delta}
.\
\end{eqnarray}
The last two terms represent the contribution of the mesons $f$s, $\phi$s and $\delta$s and read
\begin{eqnarray}
\mathcal{L}^{YY}&=&\sum_{B}g_{f_{0} B}\overline{\Psi}_{B}\Psi_{B}f_{0}-\sum_{B}g_{\phi B}\overline{\Psi}_{B}\gamma_{\mu}\Psi_{B}\phi^{\mu}
\nonumber\\
&&+\frac{1}{2}\left(\partial_{\mu}f_{0}\partial^{\mu}f_{0}-m_{f_{0}}^{2}f_{0}^{2}\right)\nonumber\\
&&-\frac{1}{4}S_{\mu \nu}S^{\mu \nu}+\frac{1}{2}m_{\phi}^{2}\phi_{\mu}\phi^{\mu}
.\
\end{eqnarray}
and
\begin{eqnarray}
\mathcal{L^{\delta}}&=&
\sum_{B}\overline{\Psi}_{B}g_{\delta B}\tau\cdot\delta\Psi_{B}
+\frac{1}{2}\partial_{\mu}\delta\partial^{\mu}\delta-\frac{1}{2}m_{\delta}^{2}\delta^{2}
.\
\end{eqnarray}
Here, $S_{\mu \nu}=\partial_{\mu}\phi_{\nu}-\partial_{\nu}\phi_{\mu}$.

For the condition of $\beta$ equilibrium, the chemical equilibrium is
\begin{eqnarray}
\mu_{i}=b_{i}\mu_{n}-q_{i}\mu_{e},
\end{eqnarray}
where $b_{i}$ is the baryon number of a species $i$.

After the RMF approach is used, the energy density $\varepsilon$ and pressure $p$ of a PNS are given by~\cite{GlendenningPlb87,Glendenningnpa87}

\begin{eqnarray}
\varepsilon&=&\frac{1}{3}g_{2}\sigma^{3}+\frac{1}{4}g_{3}\sigma^{4}+\frac{1}{2}
m_{\sigma}^{2}\sigma^{2}+\frac{1}{2}m_{\omega}^{2}\omega_{0}^{2}+\frac{1}{2}m_{\rho}^{2}\rho_{03}^{2}
\nonumber\\
&&+\sum_{B}\frac{2J_{B}+1}{2\pi^{2}}\int_{0}^{\infty}\kappa^{2}\mathrm d\kappa\sqrt{\kappa^{2}+m^{*2}}
\nonumber\\
&&\times(\mathrm e\mathrm x\mathrm p[(\varepsilon_{B}(k)-\mu_{B})/T]+1)^{-1}
\nonumber\\
&&+\sum_{\lambda=e,\mu}\frac{2J_{\lambda}+1}{2\pi^{2}}\int_{0}^{\infty}\kappa^{2}\mathrm d\kappa\sqrt{\kappa^{2}+m_{\lambda}^{2}}
\nonumber\\
&&\times(\mathrm e\mathrm x\mathrm p[(\varepsilon_{\lambda}(k)-\mu_{\lambda})/T]+1)^{-1}
\nonumber\\
&&+\frac{1}{2}m_{f}^{2}f_{0}^{2}+\frac{1}{2}m_{\phi}^{2}\phi_{0}^{2}+\frac{1}{2}m_{\delta}^{2}\delta _{0}^{2}
,\\
p&=&-\frac{1}{2}m_{\sigma}^{2}\sigma^{2}-\frac{1}{3}g_{2}\sigma^{3}-\frac{1}{4}g_{3}
+\frac{1}{2}m_{\rho}^{2}\rho_{03}^{2}
\sigma^{4}+\frac{1}{2}m_{\omega}^{2}\omega_{0}^{2}
\nonumber\\
&&+\frac{1}{3}\sum_{B}\frac{2J_{B}+1}{2\pi^{2}}\int_{0}^{\infty}\frac{\kappa^{4}}{\sqrt{\kappa^{2}+m^{*2}}}\mathrm d\kappa
\nonumber\\
&&\times(\mathrm e\mathrm x\mathrm p[(\varepsilon_{B}(k)-\mu_{B})/T]+1)^{-1}
\nonumber\\
&&+\frac{1}{3}\sum_{\lambda=e,\mu}\frac{2J_{\lambda}+1}{2\pi^{2}}\int_{0}^{\infty}\frac{\kappa^{4}}{\sqrt{\kappa^{2}+m_{\lambda}^{2}}}\mathrm d\kappa
\nonumber\\
&&\times(\mathrm e\mathrm x\mathrm p[(\varepsilon_{\lambda}(k)-\mu_{\lambda})/T]+1)^{-1}
\nonumber\\
&&-\frac{1}{2}m_{f}^{2}f_{0}^{2}+\frac{1}{2}m_{\phi}^{2}\phi_{0}^{2}-\frac{1}{2}m_{\delta}^{2}\delta _{0}^{2}
.
\end{eqnarray}
where, $m^{*}$ is the effective mass of baryons
\begin{eqnarray}
m^{*}=m_{B}-g_{\sigma B}\sigma.
\end{eqnarray}

The mass and the radius of a PNS can be calculated through the Tolman-Oppenheimer-Volkoff (TOV) equations~\cite{Glen97,Tolman39,Oppenheimer39}

\begin{eqnarray}
\frac{\mathrm dp}{\mathrm dr}&=&-\frac{\left(p+\varepsilon\right)\left(M+4\pi r^{3}p\right)}{r \left(r-2M \right)}
,\\\
M&=&4\pi\int_{0}^{{\color{blue}R}}\varepsilon r^{2}\mathrm dr
.\
\end{eqnarray}

The moment of inertia of a slowly rotating PNS is given by~\cite{Glen97}
\begin{eqnarray}
I=\frac{8\pi}{3}\int_{0}^{R}\mathrm drr^{4}\frac{\varepsilon+p}{\sqrt{1-2M(r)/r}}\frac{\left[\Omega-\omega(r)\right]}{\Omega}e^{-\nu}
.\
\end{eqnarray}
Here, $\nu$ is given by
\begin{eqnarray}
-\frac{\mathrm d\nu(r)}{\mathrm dr}=\frac{1}{\varepsilon+p}\frac{\mathrm dp}{\mathrm dr}
,\
\end{eqnarray}
and the angular velocity is given by
\begin{eqnarray}
-\frac{1}{r^{4}}\frac{\mathrm d}{\mathrm dr}\left(r^{4}j\frac{\mathrm d\overline{\omega}}{\mathrm dr}\right)
+\frac{4}{r}\frac{\mathrm dj}{\mathrm dr}\overline{\omega}=0
.\
\end{eqnarray}

The $j(r)$ is
\begin{eqnarray}
j(r)=e^{-\left(\nu+\lambda\right)}=e^{-\nu}\sqrt{1-2M(r)/r},          r<R
.\
\end{eqnarray}

The boundary condition are given by
\begin{eqnarray}
\frac{\mathrm d\overline{\omega}}{\mathrm dr}_{\mid r=0}=0
,\\
\nu\left(\infty\right)=0
,\\
\overline{\omega}\left(R\right)=\Omega-\frac{R}{3}\frac{\mathrm d\overline{\omega}}{\mathrm dr}_{\mid r=R}
.\
\end{eqnarray}

\section{The parameters}
Because larger compression modulus $K$ will give the larger maximum mass of the NS~\cite{zhaoappb12}, the nucleon coupling constant GM1, which compression modulus $K$ is larger, is chosen in this work~\cite{Glen91}: the saturation density $\rho_{0}$=0.153 fm$^{-3}$, binding energy B/A=16.3 MeV, a compression modulus $K=300$ MeV, charge symmetry coefficient $a_{sym}$=32.5 MeV and the effective mass $m^{*}/m$=0.7.

We define the ratios of hyperon coupling constant to nucleon coupling constant: $x_{\sigma h}=\frac{g_{\sigma h}}{g_{\sigma}}=x_{\sigma}$, $x_{\omega h}=\frac{g_{\omega h}}{g_{\omega}}=x_{\omega}$, $x_{\rho h}=\frac{g_{\rho h}}{g_{\rho}}$, with $h$ denoting hyperons $\Lambda, \Sigma$ and $\Xi$.

The calculations show that the ratio of hyperon coupling constant to nucleon coupling constant is in the range 1/3$\sim$1~\cite{Glen91}. The maximum mass of the NS calculated will increase with the increase of the parameter $x_{\omega h}$ and so we choose $x_{\omega h}$=0.9, .08, 0.7, ..., at first. The corresponding $x_{\sigma h}$ can be obtained by the hyperon well depth~\cite{Glen97}

\begin{eqnarray}
U_{h}^{(N)}=m_{n}\left(\frac{m_{n}^{*}}{m_{n}}-1\right)x_{\sigma h}+\left(\frac{g_{\omega}}{m_{\omega}}\right)^{2}\rho_{0}x_{\omega h}
.\
\end{eqnarray}
Here, we choose $U_{\Lambda}^{(N)}=-30$ MeV~\cite{Schaff00}, $ U_{\Sigma}^{(N)}=+20$ MeV~\cite{Kohno10} and $U_{\Xi}^{(N)}=-14$ MeV~\cite{{Harada10}}. For the $x_{\rho h}$, we select $x_{\rho \Lambda}=0$, $x_{\rho \Sigma}=2$, $x_{\rho \Xi}=1$ by SU(6) symmetry~\cite{Schaff96}.

For the coupling constants of the mesons $f$ and $\phi$, we choose as~\cite{Schaffner94}
\begin{eqnarray}
g_{f \Lambda}/g_{\sigma}=g_{f \Sigma}/g_{\sigma}=0.69, g_{f \Xi}/g_{\sigma}=1.25
,\\
g_{\phi \Xi}=2g_{\phi \Lambda}=2g_{\phi \Sigma}=-2\sqrt{2}g_{\omega}/3
.\
\end{eqnarray}
The coupling constants of mesons $\delta$ are chosen as~\cite{Menezes04}
\begin{eqnarray}
g_{\delta \Sigma }=2g_{\delta \Xi }=2g_{\delta}, g_{\delta \Lambda}=0.
\end{eqnarray}
Here, the temperature is chosen as $T=20$ MeV~\cite{Burrows86}.

For the mesons $f, \phi$ and $\delta$ will soften the equation of state (EoS) and the softer EoS will give smaller PNS mass. Therefore, we calculate the mass of the PNS containing mesons $f, \phi$ and $\delta$ at first.

\begin{figure}[!htp]
\centering{}\includegraphics[width=4.5in]{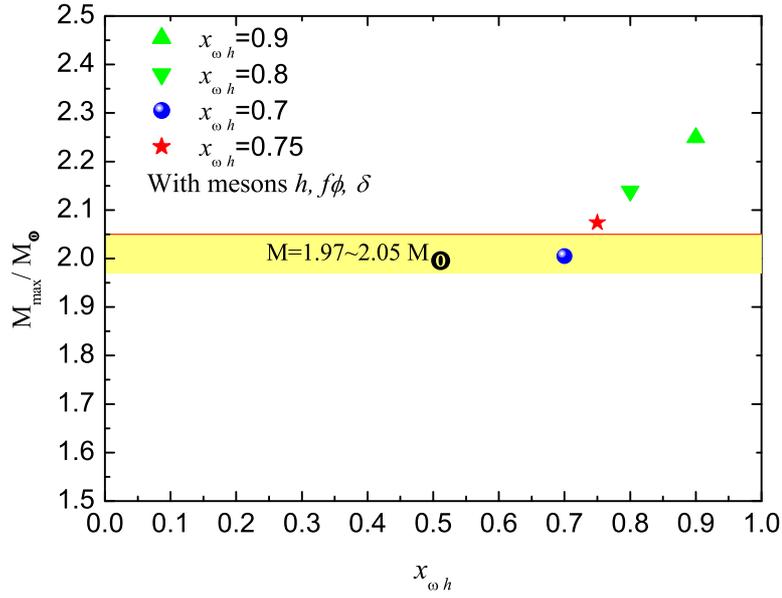}\caption{The maximum mass $M_{max}$ of the PNS as a function of the parameter $x_{\omega h}$.}
\label{fig1}
\end{figure}

For parameters $x_{\omega h}$=0.9, 0.8, we get two parameter sets: No.1 ($x_{\sigma \Lambda}$=0.7961, $x_{\omega \Lambda}$=0.9; $x_{\sigma \Sigma}$=0.6186, $x_{\omega \Sigma}$=0.9; $x_{\sigma \Xi}$=0.7393, $x_{\omega \Xi}$=0.9) and No.2 ($x_{\sigma \Lambda}$=0.7194, $x_{\omega \Lambda}$=0.8; $x_{\sigma \Sigma}$=0.5420, $x_{\omega \Sigma}$=0.8; $x_{\sigma \Xi}$=0.6627, $x_{\omega \Xi}$=0.8), by which the maximum mass obtained are $M_{max}$=2.249, 2.139 M$\odot$, respectively (see Fig.~\ref{fig1}). We hope to calculate the maximum mass of PNS as close as possible to the mass of the PNS PSR J0348+0432, where the influence of the mesons will be more obvious. But parameters No.01 and No.02 give the too larger maximum mass. So, we continue to select No.3 ($x_{\sigma \Lambda}$=0.6428, $x_{\omega \Lambda}$=0.7; $x_{\sigma \Sigma}$=0.4653, $x_{\omega \Sigma}$=0.7; $x_{\sigma \Xi}$=0.5860, $x_{\omega \Xi}$=0.7), by which the maximum mass of the PNS calculated only is 2.005 M$_{\odot}$, still smaller than the mass of the PNS PSR J0348+0432. Then, we further choose No.4 ($x_{\sigma \Lambda}$=0.6812, $x_{\omega \Lambda}$=0.75; $x_{\sigma \Sigma}$=0.5037, $x_{\omega \Sigma}$=0.75; $x_{\sigma \Xi}$=0.6244, $x_{\omega \Xi}$=0.75) and in this case we obtain the maximum mass $M_{max}$=2.074 M$_{\odot}$, which is close to the mass of the PNS PSR J0348+0432.

Next, we use parameter No.04 to study the effect of $f$s, $\phi$s and $\delta$s on the moment of inertia of the PNS PSR J0348+0432.

\section{Effect of mesons $f, \phi$ and $\delta$ on energy density and pressure of the PNS PSR J0348+0432}
\begin{figure}[!htp]
\centering{}\includegraphics[width=4.5in]{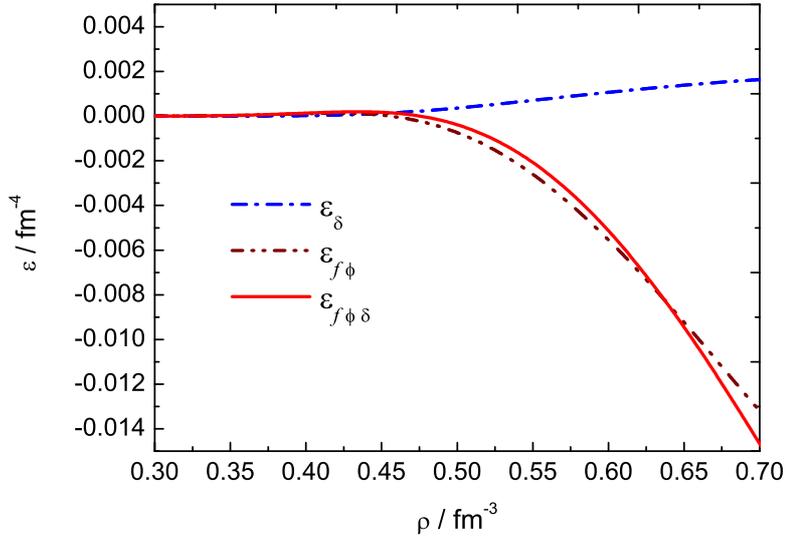}\caption{The contribution of mesons $f$, $\phi$ and $\delta$ to energy density $\varepsilon_{\delta}$, $\varepsilon_{f \phi}$ and $\varepsilon_{f \phi \delta}$ as a function of the baryon density $\rho$. Here, $x_{\omega h}$=0.75.}
\label{fig2}
\end{figure}

The contribution of mesons $f$, $\phi$ and $\delta$ to energy density $\varepsilon_{\delta}$, $\varepsilon_{f \phi}$ and $\varepsilon_{f \phi \delta}$ as a function of the baryon density $\rho$ are shown in Fig.~\ref{fig2} and Table~\ref{tab1}. Here, $x_{\omega h}$=0.75.

\begin{table}[!htbp]
\centering
\caption{The central particle number density $\rho_{c}$, the central energy density $\varepsilon_{c}$, $\varepsilon_{c, \delta}$, $\varepsilon_{c, f \phi}$, $\varepsilon_{c, f \phi \delta}$ and the central pressure $p_{c}$, $p_{c, \delta}$, $p_{c, f \phi}$ and $p_{c, f \phi \delta}$ of the PNS PSR J0348+0432 $M$. Here, $x_{\omega h}$=0.75.
The $\varepsilon_{c}$, $\varepsilon_{c, \delta}$, $\varepsilon_{c, f \phi}$, $\varepsilon_{c, f \phi \delta}$, $p_{c}$, $p_{c, \delta}$, $p_{c, f \phi}$ and $p_{c, f \phi \delta}$ are in the units of fm$^{-4}$, $\rho_{c}$ fm$^{-3}$, }
\label{tab1}
\begin{tabular}[t]{ccccc}
\hline\noalign{\smallskip}
Quantity         &$h$       &$h \delta$&$h f \phi$&$h f \phi \delta$\\
\hline
$\rho_{c}$       &0.526$\sim$0.618&0.526$\sim$0.613&0.542$\sim$0.665&0.544$\sim$0.685\\
$\varepsilon_{c}$&2.857$\sim$3.473&2.858$\sim$3.446&2.960$\sim$3.789&2.974$\sim$3.928\\
$p_{c}$&0.575$\sim$0.780&0.575$\sim$0.775&0.590$\sim$0.865&0.590$\sim$0.900\\
\hline\noalign{\smallskip}
\hline
Quantity&$h$       &$\delta$&$f \phi$&$f \phi \delta$\\
\hline
$\varepsilon_{c, \delta}$&&5.339E-4$\sim$1.15E-3&&\\
$\varepsilon_{c, f \phi}$&&&-2.23E-3$\sim$-1.041E-2&\\
$\varepsilon_{c, f \phi \delta}$&&&&-1.8E-3$\sim$-1.304E-2\\
$p_{c, \delta}$&&1.09E-3$\sim$7.66E-3&&\\
$p_{c, f \phi}$&&&-1.831E-2$\sim$-3.225E-2&\\
$p_{c, f \phi \delta}$&&&&-2.279E-2$\sim$-4.968E-2\\
\noalign{\smallskip}\hline\noalign{\smallskip}
\end{tabular}
\vspace*{0.6cm}  
\end{table}

From Table~\ref{tab1} we see that the central particle number density of the PNS PSR J0348+0432 is in the range $\rho_{c}$=0.526$\sim$0.618 fm$^{-3}$ as mesons $f$, $\phi$ and $\delta$ not being considered. When the mesons $\delta$ are included, the central particle number density decreases to $\rho_{c}$=0.526$\sim$0.613 fm$^{-3}$ whereas it will increase to $\rho_{c}$=0.542$\sim$0.665 fm$^{-3}$ as the mesons $f$ and $\phi$ being considered. Thereafter, the central particle number density will increase to $\rho_{c}$=0.544$\sim$0.685 fm$^{-3}$ when the mesons $f$, $\phi$ and $\delta$ are all considered.

We see that the contribution of mesons $\delta$ to energy density $\varepsilon_{\delta}$ is greater than zero while the contribution of mesons $f$ and $\phi$ to energy density $\varepsilon_{f \phi}$ is less than zero. These mean that the energy density $\varepsilon$ will increase considering the mesons $\delta$ whereas it will decrease as the mesons $f$ and $\phi$ being considered. The magnitude of the decline is greater than the magnitude of the increase. Therefore, the contribution of the mesons $f$, $\phi$ and $\delta$ to the energy density $\varepsilon_{f \phi \delta}$ will be less than zero (see Fig.~\ref{fig2}).

\begin{figure}[!htp]
\centering{}\includegraphics[width=4.5in]{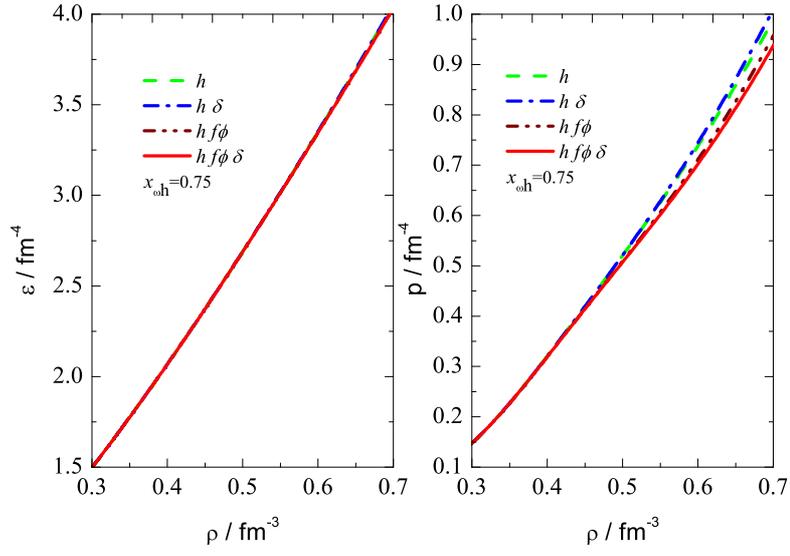}\caption{The energy density $\varepsilon$ and pressure $p$ as a function of the baryon density $\rho$.}
\label{fig3}
\end{figure}

From Table~\ref{tab1} we also see the ratio of various central energy density is $|\varepsilon_{c}|$: $|\varepsilon_{c, f \phi \delta}|$ : $|\varepsilon_{c, f \phi}|$: $|\varepsilon_{c, \delta}|$ $\cong$ 5000$\sim$6500: 3$\sim$24: 4$\sim$20: 1$\sim$2. We see that the contribution of mesons $f$ and $\phi$ to central energy density is 4$\sim$10 times larger than that of mesons $\delta$. But the contribution of mesons $f$, $\phi$ and $\delta$ to the energy density is still very small, only the energy density's 0.06$\sim$0.6\% (see Fig.~\ref{fig3}).

\begin{figure}[!htp]
\centering{}\includegraphics[width=4.5in]{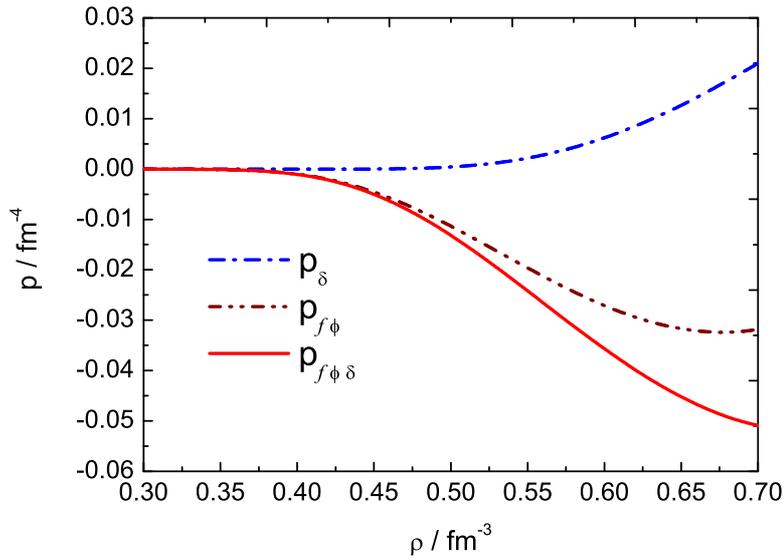}\caption{The central pressure $p_{c, \delta}$, $p_{c, f \phi}$ and $p_{c, f \phi \delta}$ of the PNS PSR J0348+0432 as a function of the baryon density $\rho$.}
\label{fig4}
\end{figure}

The central pressure $p_{c, \delta}$, $p_{c, f \phi}$ and $p_{c, f \phi \delta}$ of the PNS PSR J0348+0432 as a function of the baryon density $\rho$ are shown in Fig~\ref{fig4} (also see Table.~\ref{tab1}).

We see that the contribution of the mesons $\delta$ to pressure $p_{\delta}$ is greater than zero while the contribution of the mesons $f$ and $\phi$ to pressure $p_{f \phi}$ is less than zero. These mean that the pressure will increase considering the mesons $\delta$ and  will decrease as the mesons $f$ and $\phi$ being considered. The magnitude of the decline is greater than that of the increase. Therefore, the contribution of the mesons $f$, $\phi$ and $\delta$ to the pressure $p_{f \phi \delta}$ will be less than zero (see Fig.~\ref{fig4}).

We also see from Table~\ref{tab1} that the ratio of various central pressure is $|p_{c}|$: $|p_{c, f \phi \delta}|$ : $|p_{c, f \phi}|$: $|p_{c, \delta}|$ $\cong$ 500$\sim$700: 20$\sim$50: 20$\sim$30: 1$\sim$7. We see that the contribution of mesons $f$ and $\phi$ to central pressure is 20$\sim$4 times larger than that of mesons $\delta$. But the contribution of mesons $f$, $\phi$ and $\delta$ to the central pressure is still very small, only the central pressure's 4$\sim$7\% (see Fig.~\ref{fig3}).

\section{Effect of the mesons $f, \phi$ and $\delta$ on the radius $R$ of the PNS PSR J0348+0432}
\begin{figure}[!htp]
\centering{}\includegraphics[width=4.5in]{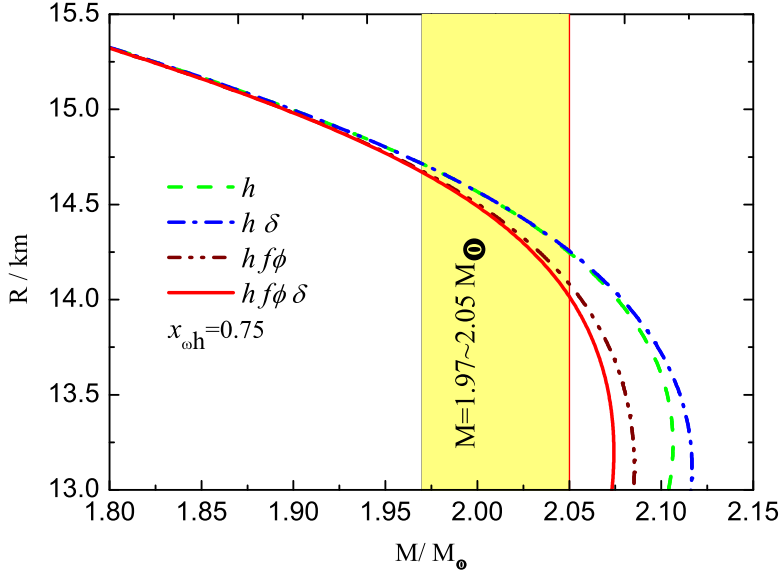}\caption{The radius $R$ as a function of the mass $M$.}
\label{fig5}
\end{figure}

Figure~\ref{fig5} gives the radius $R$ as a function of the mass $M$ (also see Table~\ref{tab2}). We see the radius increase as the mesons $\delta$ being considered whereas decrease when the mesons $f$ and $\phi$ included. Because the magnitude of the decline is larger than that of increase, the radius will decrease considered the contribution of mesons $f$, $\phi$ and $\delta$.

\begin{table}[!htbp]
\centering
\caption{The radius $R$, the moment of inertia of the PNS PSR J0348+0432 $I$ and the contribution of mesons $f$, $\phi$ and $\delta$ to the moment of inertia $I_{\delta}$, $I_{f \phi}$ and $I_{f \phi \delta}$. Here, $x_{\omega h}$=0.75.
The $R$ is in the unit km, $I$, $I_{\delta}$, $I_{f \phi}$ and $I_{f \phi \delta}$ $\times$10$^{45}$ g cm$^{2}$.}
\label{tab2}
\begin{tabular}[t]{ccccc}
\hline\noalign{\smallskip}
Quantity       &$h$           &$h \delta$    &$h f\phi$       &$h \delta f\phi$        \\
\hline
$R$           &14.708$\sim$14.246&14.708$\sim$14.255&14.670$\sim$14.080&14.671$\sim$14.108       \\
$I$           &2.286$\sim$2.149&2.287$\sim$2.161&2.241$\sim$2.044&2.233$\sim$2.003       \\
\hline\noalign{\smallskip}
Quantity       &$h$           &$\delta$    &$f\phi$       &$\delta f\phi$        \\
\hline
$I_{\delta}$           &&1.224E42$\sim$1.161E43&&       \\
$I_{f \phi}$           &&&-4.541E43$\sim$-1.057E44&       \\
$I_{f \phi \delta}$           &&&&-5.302E43$\sim$-1.465E44       \\
\noalign{\smallskip}\hline\noalign{\smallskip}
\end{tabular}
\vspace*{0.6cm}  
\end{table}

Corresponding to the PNS PSR J0348+0432 ($M$=1.97$\sim$2.05 M$_{\odot}$), the radius is in the range $R=14.708\sim14.246$ km as mesons $f$, $\phi$ and $\delta$ not being considered. When we consider mesons $\delta$ the radius increases to $R=14.708\sim14.255$ km whereas decreases to $R=14.670\sim14.080$ km as mesons $f$ and $\phi$ being considered. Thus, as mesons $f$, $\phi$ and $\delta$ all being considered the radius will decrease to $R=14.671\sim14.108$ km.

\section{Effect of the mesons $f, \phi$ and $\delta$ on the moment of inertia of the PNS PSR J0348+0432}
\begin{figure}[!htp]
\centering{}\includegraphics[width=4.5in]{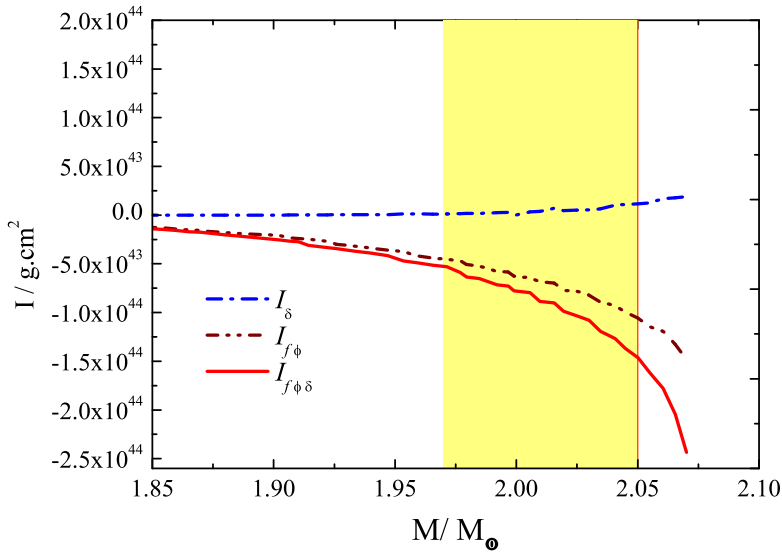}\caption{The contribution of mesons $f$, $\phi$ and $\delta$ to the moment of inertia $I_{\delta}$, $I_{f \phi}$ and $I_{f \phi \delta}$ as a function of the mass $M$.}
\label{fig6}
\end{figure}

The contribution of mesons $f$, $\phi$ and $\delta$ to the moment of inertia $I_{\delta}$, $I_{f \phi}$ and $I_{f \phi \delta}$ as a function of the mass $M$ are shown in Fig.~\ref{fig6} and Table~\ref{tab2}.

From Fig~\ref{fig6} we see that the contribution of mesons $\delta$ to moment of inertia $I_{\delta}$ is greater than zero whereas the contribution of mesons $f$ and $\phi$ to moment of inertia $I_{f \phi}$ is less than zero. That is to say that the moment of inertia $I$ will increase considering the mesons $\delta$ and will decrease as the mesons $f$ and $\phi$ being considered. For the magnitude of the decline is greater than the magnitude of the increase and therefore the contribution of mesons $f$, $\phi$ and $\delta$ to the moment of inertia $I_{f \phi \delta}$ will be less than zero (see Fig.~\ref{fig6}).

\begin{figure}[!htp]
\centering{}\includegraphics[width=4.5in]{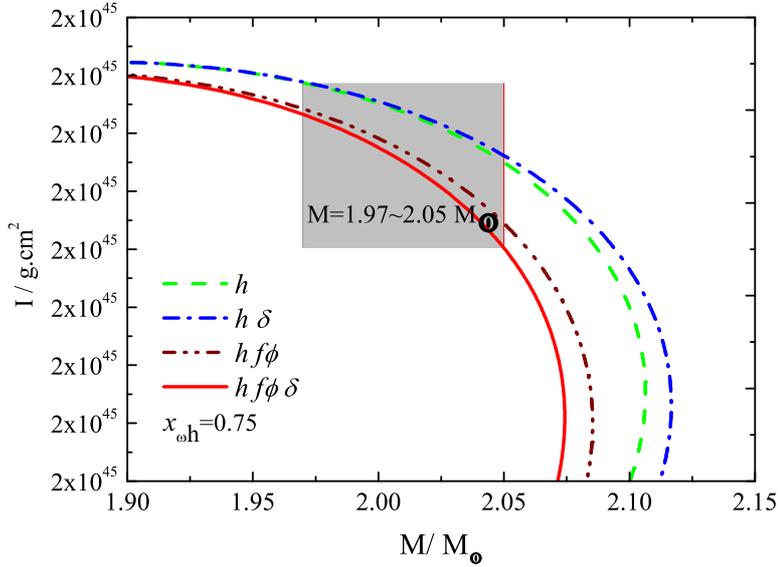}\caption{The moment of inertia $I$ as a function of the mass $M$.}
\label{fig7}
\end{figure}

From Table~\ref{tab2} we also see the ratio of various moment of inertia is $|I|$: $|I_{f \phi \delta}|$ : $|I_{f \phi}|$: $|I_{\delta}|$ $\cong$ 1900$\sim$1800: 40$\sim$120: 40$\sim$90: 1$\sim$10. We see that the contribution of mesons $f$ and $\phi$ to moment of inertia is 4$\sim$9 times larger than that of mesons $\delta$. We also see that the contribution of mesons $f$, $\phi$ and $\delta$ to the moment of inertia is still very small, only the moment of inertia's 2$\sim$5\% (see Fig.~\ref{fig7}).

\section{Summary}
In this paper, we calculate the effect of $f_{0}(975)$s, $\phi(1020)$s and $\delta$s on the moment of inertia of the PNS PSR J0348+0432 in the framework of the RMF theory considering the baryon octet. In this work, we choose the nucleon coupling constants GM1, the hyperon well depth $U_{\Lambda}^{(N)}$=-30 MeV, $U_{\Sigma}^{(N)}$=+20 MeV, $U_{\Xi}^{(N)}$=-14 MeV and the temperature $T$=20 MeV.

We find that the contributions of mesons $\delta$ to energy density $\varepsilon_{\delta}$ and pressure $p_{\delta}$ are greater than zero while the contributions of mesons $f$ and $\phi$ to energy density $\varepsilon_{f \phi}$ and pressure $p_{f \phi}$ are less than zero. These cause the energy density $\varepsilon$ and pressure $p$ increase considering the mesons $\delta$ whereas will decrease as the mesons $f \phi$ being considered. Because of the magnitude of the decline being greater than the magnitude of the increase, the $\varepsilon_{f \phi \delta}$ will decrease considering the mesons $f, \phi$ and $\delta$.

The contribution of mesons $f$ and $\phi$ to central energy density is 4$\sim$10 times larger than that of mesons $\delta$ and the contribution of mesons $f$ and $\phi$ to central pressure is 20$\sim$4 times larger than that of mesons $\delta$. We also see that the contribution of mesons $f$, $\phi$ and $\delta$ to the central energy density is only the energy density's 0.06$\sim$0.6\% but the contribution of mesons $f$, $\phi$ and $\delta$ to the central pressure is the central pressure's 4$\sim$7\%.

The radius increase as the mesons $\delta$ being considered whereas decrease as the mesons $f$ and $\phi$ being included. The magnitude of the decline being larger than that of the increase leads to the radius decrease when the contributions of mesons $f$, $\phi$ and $\delta$ are considered.

Similarly, the contribution of mesons $\delta$ to moment of inertia $I_{\delta}$ is greater than zero whereas the contribution of mesons $f$ and $\phi$ to moment of inertia $I_{f \phi}$ is less than zero, i.e. the moment of inertia $I$ will increase considering the mesons $\delta$ whereas will decrease as the mesons $f$ and $\phi$ being considered. As a result of the magnitude of the decline being greater than the magnitude of the increase the $I_{f \phi \delta}$ will decrease considering the mesons $f$, $\phi$ and $\delta$. In addition, we also find that the contribution of mesons $f$ and $\phi$ to moment of inertia is 4$\sim$9 times larger than that of mesons $\delta$ and the mesons $f$, $\phi$ and $\delta$ contribute to the moment of inertia's 2$\sim$5\%.

\begin{acknowledgements}
We are thankful to Shan-Gui Zhou for fruitful discussions during my visit to the Institute of Theoretical Physics of Chinese Academy of Sciences.
This work was supported by the Natural Science Foundation of China ( Grant No. 11447003 ) and the Scientific Research Foundation of the Higher Education Institutions of Anhui Province, China ( Grant No. KJ2014A182 ).
\end{acknowledgements}


\end{document}